\documentclass[iop]{emulateapj}

\usepackage{color,eps fig,graphics}

\newcommand{\eg}{e.g.,}

\newcommand{\etal}{{et al.}}
\newcommand{\kms}{{km~s$^{-1}$}}
\newcommand{\sersic}{S\'{e}rsic}

\newcommand{\msun}{$M_{\odot}$}

\begin{document}

\title{A SEARCH FOR MODERATE-REDSHIFT SURVIVORS FROM THE POPULATION OF LUMINOUS COMPACT PASSIVE GALAXIES AT HIGH REDSHIFT\altaffilmark{1}}
\author{Alan Stockton, Hsin-Yi Shih, Kirsten Larson, and Andrew W. Mann\altaffilmark{2}}
\affil{Institute for Astronomy, University of Hawaii, Honolulu, HI 96822; stockton@ifa.hawaii.edu, hsshih@ifa.hawaii.edu, klarson@ifa.hawaii.edu, amann@ifa.hawaii.edu}

\altaffiltext{1}{Some of the data presented herein were obtained at the W.M. Keck Observatory, which is operated as a scientific partnership among the California Institute of Technology, the University of California and the National Aeronautics and Space Administration. The Observatory was made possible by the generous financial support of the W.M. Keck Foundation.
}
\altaffiltext{2}{Now at Department of Astronomy, The University of Texas at Austin, Austin, TX 78712}

\begin{abstract}
From a search of a $\sim2400$ deg$^2$ region covered by both the Sloan Digital Sky Survey and UKIRT Infrared Deep Sky Survey databases, we have attempted to identify galaxies at $z\sim0.5$ that are consistent with their being essentially unmodified examples of the luminous passive compact galaxies found at $z\sim2.5$. After isolating good candidates via deeper imaging, we further refine the sample with Keck moderate-resolution spectroscopy and laser guide star adaptive-optics imaging. For four of the five galaxies that so far remain after passing through this sieve, we analyze plausible star-formation histories based on our spectra in order to identify galaxies that may have survived with little modification from the population formed at high redshift. We find two galaxies that are consistent with having formed $\gtrsim95$ \%\ of their mass at $z>5$. We attempt to estimate masses both from our stellar population determinations and from velocity dispersions. Given the high frequency of small axial ratios, both in our small sample and among samples found at high redshifts, we tentatively suggest that some of the more extreme examples of passive compact galaxies may have prolate morphologies.
\end{abstract}

\keywords{galaxies: formation,---galaxies: kinematics and dynamics,---galaxies: stellar content,---galaxies: structure}

\section{Introduction}

Massive galaxies found at $z\sim2.5$ comprising only relatively old ($\sim2$ Gyr) stars appear to be essentially pristine examples of galaxies that apparently formed $\gtrsim10^{11} M_{\sun}$ at redshifts of 5--10 or more. Virtually all of these, as well as a significant fraction of the population of passive galaxies at $z\sim1.5$, are extremely compact when compared with galaxies of similar mass in the present-day universe \citep[\eg][]{sto04, dad05, tru07, zir07, tof07, mcg08, vdok08, bui08, dam09, muz09}.
Importantly, these massive compact galaxies must each have formed in a strongly dissipational event over a very short time period; they are remnants of the very first major episodes of star formation in the universe.
The morphologies of these galaxies are important because they can preserve information relevant to formation mechanisms of early-epoch massive galaxies in general. \citet{kri06} found that $\sim45$\% of the massive $K$-band-selected galaxies at $z \sim 2.3$ have old stellar populations, and \citet{kri08} estimated that $\sim15$\% of the stellar mass in red-sequence galaxies with masses above $10^{11} M_{\odot}$ today was already in place in similar-mass galaxies on the red sequence at $z\sim2.3$. In any case, a significant fraction of the {\it very oldest} stars incorporated into the most massive present-day galaxies may have had their origin in these passive compact galaxies at high redshifts. \citet{vdok08}, using high resolution Hubble Space Telescope ({\it HST}) images of the Kriek et al. sample, estimated that $\sim 90$\%--100\% of these massive galaxies with old stellar populations are extremely compact, with mean effective radii of $< 1$ kpc. In Figure~1, we show the most extreme example of such a galaxy that we have found in a survey of $z\sim2.5$ radio-source fields (A.~Stockton \etal, in preparation). 
\begin{figure}[!t]
\epsscale{1.0}
\plotone{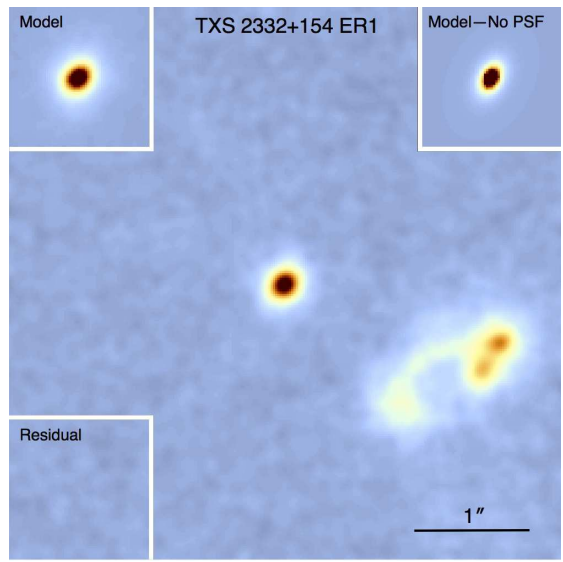}
\caption{Image of the $z=2.472$ passive compact galaxy TXS\,2332+154 ER1, obtained with the laser guide star adaptive-optics (LGSAO) system and the NIRC2 camera on the Keck II telescope. The upper-left inset is the {\sc Galfit} \citep{pen02,pen10} model, the lower-left inset shows the residual after subtraction of the model from the observed image, and the upper-right inset shows the model without convolution with the PSF.}\label{txs2332}
\end{figure}
This galaxy, at a redshift of 2.472, has an apparent stellar mass (based on a model fit) of $\sim3.5\times10^{11}$ $M_{\odot}$, a stellar population age of 1.9 Gyr for an exponentially declining star-formation model with an $e$-folding time of 0.3 Gyr, a \sersic\ index $n=3.2$, and a circularized effective radius of 360 pc.

These massive compact galaxies with old stellar populations at high redshifts have posed some real enigmas:  (1) several of these galaxies for which good photometry exists appear to have reasonably well-determined stellar population ages.\footnotemark\footnotetext{There are of course all sorts of caveats regarding ages of integrated stellar populations. Here we refer to ages determined from the best-fitting \citet{bru03} models, which even now appear to be among the more realistic spectral-synthesis models currently available \citep[\eg][]{zib12}. We have explored instantaneous burst and exponentially decreasing star-formation models with 0.4, 1.0, and 2.5 solar metallicities. For galaxies at $z\sim2.5$, the universe is only $\sim2.6$ Gyr old, integrated spectra of passive galaxies change rapidly, and the ages are unlikely to be too far off. For lower-redshift galaxies, absolute ages are more suspect, but at least {\it relative} ages should be roughly reliable.} These indicate that, for at least some massive galaxies, the bulk of the stellar population was in place by $z\sim10$. Such an early epoch of star formation would imply an average star formation rate of $\sim1000\ M_{\odot}$ per year sustained for a few $\times10^8$ yr, or even higher rates for shorter periods. (2) The very high stellar density implied by the mass and compactness of these galaxies requires an extreme degree of dissipation, {probably coupled with extreme starburst and/or quasar superwinds.} The only detailed mechanisms proposed so far (merging extremely gas-rich galaxies; \citealt{wuy10,som10}) predict that the compact galaxies should be surrounded by a faint extended distribution of old stars and have high S\'{e}rsic indices, neither of which is consistently observed. Extremely deep {\it HST} Wide-Field Camera 3 (WFC3) images \citep[\eg][]{cas10} show no evidence for the predicted extended envelopes, and some of the best studied high-redshift compact galaxies show low-to-moderate \sersic\ indices \citep{sto04,sto07,vdok08}. 
(3)  What is the final fate of these galaxies, and why are they so extremely rare at the present epoch? One possibility that has been put forward is the so-called ``inside-out'' buildup of present-day massive galaxies \citep[\eg][]{bez09, hop09,vdS13}. In this scenario, envelopes are acquired over time by the massive compact galaxies, which have become the cores of the most massive local galaxies. The extreme dearth of unmodified survivors in the present-day universe indicates that such an addition of material cannot be an infrequent, highly stochastic process, such as major mergers \citep{tay09}. Instead, there would have to be a more-or-less steady accretion of material, possibly via many minor mergers. However, there is disagreement whether the amount of material that can reasonably be added is sufficient to explain the necessary size growth \citep[see, e.g.,][]{ypen10,oogi13,vdS13}. Whether this lack of massive compact galaxies at the present epoch is a serious matter or not must await a more careful comparison of (comoving) volume densities between $z\sim2.5$ and $z\sim0$ than is yet available, {\it with special care to match morphological criteria and mass ranges}. {See \citet{qui13} for a discussion of this still unresolved issue.}

At high redshifts, we can determine general spectral-energy distributions (SEDs) from photometry or very low resolution spectroscopy, and we can estimate global morphologies of these galaxies from high-resolution imaging, with either adaptive optics with large ground-based telescopes or the {\it HST} WFC3. But it is extremely difficult to do much more than this with current facilities. In a heroic effort, \citet{vdok09} were able to measure an absorption feature in a galaxy at $z=2.19$, but this took 29 hr of observing time on an 8 m telescope. Very few additional galaxies at $z\sim2$ now have similar quality spectroscopy \citep{vdS11,vdS13,tof12}. Until much larger telescopes (ground- or space-based) are available, it will be extremely difficult to learn much more in detail about the luminous compact passive galaxies at $z\gtrsim2$ than we are able to determine now. In our opinion, the best way to make progress at the moment is to identify and study, at lower redshifts,  galaxies from this population that have survived relatively unscathed, or at least galaxies that appear to have had similar formation histories.

This approach, of course, has its own difficulties. First of all, our pilot studies so far have indicated that, at $z\sim0.5$, galaxies that are essentially unmodified survivors of typical examples found at high redshifts, or even close analogs to these, are extremely rare. We estimate that, for those with masses $>10^{11} M_{\odot}$, effective radius $R_{\rm e} \lesssim 1$ kpc, and $z<0.6$, the surface density on the sky of such galaxies is probably $<0.02$ deg$^{-2}$ and quite possibly much less than this. Second, we have to deal with the question of how we can know that examples that we do manage to find actually are examples of, or at least represent true analogs to, those at high redshift. We discuss these issues below in some detail, but the bottom line is that we believe that it is possible to identify a very small sample of galaxies at $z\sim0.5$ that are apparently true ``survivors'' from high redshift, having formed essentially all of their mass in stars at $z>5$; others among our candidates may have formed more recently, but have similar histories, including, critically, extreme dissipation. At $z\sim0.5$, we are not only able to explore the morphologies and color gradients of the galaxies to much lower surface brightnesses than we can at high redshift, but we can also potentially obtain spectroscopy of sufficient S/N to determine detailed properties of the stellar populations, such as kinematics and metallicities.

\section{Finding Needles in a Very Large Haystack}

There have been a number of recent efforts to identify compact passive galaxies at low redshifts. Taylor \etal\ (2009) searched the Sloan Digital Sky Survey (SDSS) database for early-type compact galaxies in the redshift range $0.066<z<0.12$, using both SDSS spectra and photometric redshifts. They found a number of galaxies with $1.1<R_{\rm e}<1.5$, but these all had indicated masses $\sim5\times10^{10}$ \msun, so they are not really comparable to the more massive compact galaxies found at high redshifts. \citet{tay09} conclude that such galaxies must be extremely rare at the present epoch and that their size evolution cannot be a result of a stochastic mechanism such as major merging.  On the other hand, \citet{val10} have claimed to have found substantial numbers of massive, old, compact galaxies in nearby X-ray-selected clusters, including a small number with $M>10^{11}$ \msun\ and $R_{\rm e} < 1.5$ kpc that may be of interest. However, the actual identifications of these galaxies were not given. \citet{tru09} have published a list of compact passive galaxies with $z<0.2$ culled from the SDSS. We have investigated three of these with deeper, higher resolution imaging \citep{shi11}. Two of the galaxies had effective radii and inferred stellar masses that agreed with those estimated from SDSS ($R_{\rm e}\sim1.4$ kpc, $M\sim1.5\times10^{11} M_{\odot}$), but the third had an $R_{\rm e}$ nearly three times larger than the SDSS estimate. Furthermore, {\it all} of these galaxies turned out to have quite complex structures for which single \sersic\ fits left quite large systematic residuals. \citet{fer12} have made detailed observations, including spectroscopy, of another seven of the galaxies identified by \citet{tru09}, finding that 
for essentially all of these galaxies, the light (and sometimes the mass) is dominated by stellar populations younger than 2 Gyr. \citet{pog13} have searched the Padova Millennium Galaxy and Group Catalogue \citep{cal11} for superdense galaxies in the redshift range $0.03<z<0.11$, covering $\sim38$ deg$^2$ of sky and finding some 32 galaxies that meet their criteria. All but three of these, however, have indicated masses $<10^{11} M_{\odot}$. Most recently, \citet{dam13} have identified nine compact galaxies at intermediate redshift, including two with essentially pure old stellar populations. However, as far as we can determine, none of these searches seem to have turned up galaxies like the more compact massive ones typically found at high redshift, with $R_{\rm e}\lesssim1$ kpc and $M>10^{11} M_{\odot}$.

The principal goals of our program have been to carry out a systematic search for minimally altered moderate-redshift examples of the luminous, passive, compact galaxies found at high redshifts, as well as to identify closely related objects that might shed more light on the subsequent evolution of these galaxies. Our primary resources for this search have been the SDSS and UKIRT Infrared Deep Sky Survey (UKIDSS; \citealt{law07}) databases. Our search has focused on higher redshifts ($z\sim0.5$) than the searches mentioned above, with the hope that at this redshift we might have a better chance of isolating at least a few massive compact galaxies that had survived from the high-redshift population relatively unscathed. On the other hand, this redshift is still sufficiently low that we can obtain high S/N spectra and LGSAO imaging of sufficient quality to allow us to explore in considerable detail morphologies, stellar populations, velocity dispersions, and other physical properties. {Our search is by no means complete, in that we have not been able to explore all of the relevant parameter space, nor have we yet been able to follow up every potential candidate. In addition, a single errant SDSS or UKIDSS magnitude outside the typical error estimates (which we find to be not terribly uncommon) will throw an object out of our sample. Nevertheless, we feel that we have done sufficient exploration to confirm that galaxies with old stellar populations that meet our criteria ($R_{\rm e}\lesssim1$ kpc and $M>10^{11} M_{\odot}$) exist at $z\sim0.5$, but we also find that they are extremely rare.}

We select candidates that have SEDs matching those of old stellar populations, where the spectral synthesis models are calculated for each interval of 0.05 in redshift. We also impose a ``compactness'' criterion based on the difference between the SDSS ``model'' and point-spread function (PSF) magnitudes; many of the objects selected by our search criteria and that finally turn out to be of interest are classified as stars by the SDSS. For objects at $z\sim0.5$, the SDSS almost never has spectra, although occasionally one will have been flagged as a high-redshift QSO candidate and have a spectrum for that reason. We indeed find occasional contamination from certain types of carbon stars and from red QSOs, but over 90\% of the objects found by our current selection procedures are galaxies at close to the expected redshift, so we no longer have found it worthwhile to obtain low S/N spectra to further refine the sample as a first step. However, we do first have to obtain deeper imaging observations at finer pixel scales, since both the SDSS and UKIDSS have 0\farcs4 pixels and do not go deep enough to show faint envelopes. This imaging program is an essential step in sifting our sample: the great majority of our original candidates turn out to have $R_{\rm e}$ much greater than 1 kpc (for best-fit single \sersic\ profiles).

\section{Five Massive Compact Galaxies at \boldmath $z\sim0.5$}\label{five}

We have carried out our search over $\sim2400$ deg$^2$ of sky. We have so far identified just five galaxies that meet our selection criteria. Two of these have been published before \citep[][Article 1]{sto10}, but we reanalyze these as well as the new ones here. (Note that Article 1 quotes major axis values for $R_{\rm e}$, whereas in this article we use circularized values, unless clearly specified otherwise, as in the following paragraph). All of the spectroscopy we show has been obtained with LRIS \citep{oke95} on the Keck I telescope, and all of the high-resolution imaging has come from the LGSAO system \citep{wiz06} and the NIRC2 camera on the Keck II telescope.

\begin{figure*}[!p]
\epsscale{1.00}
\plotone{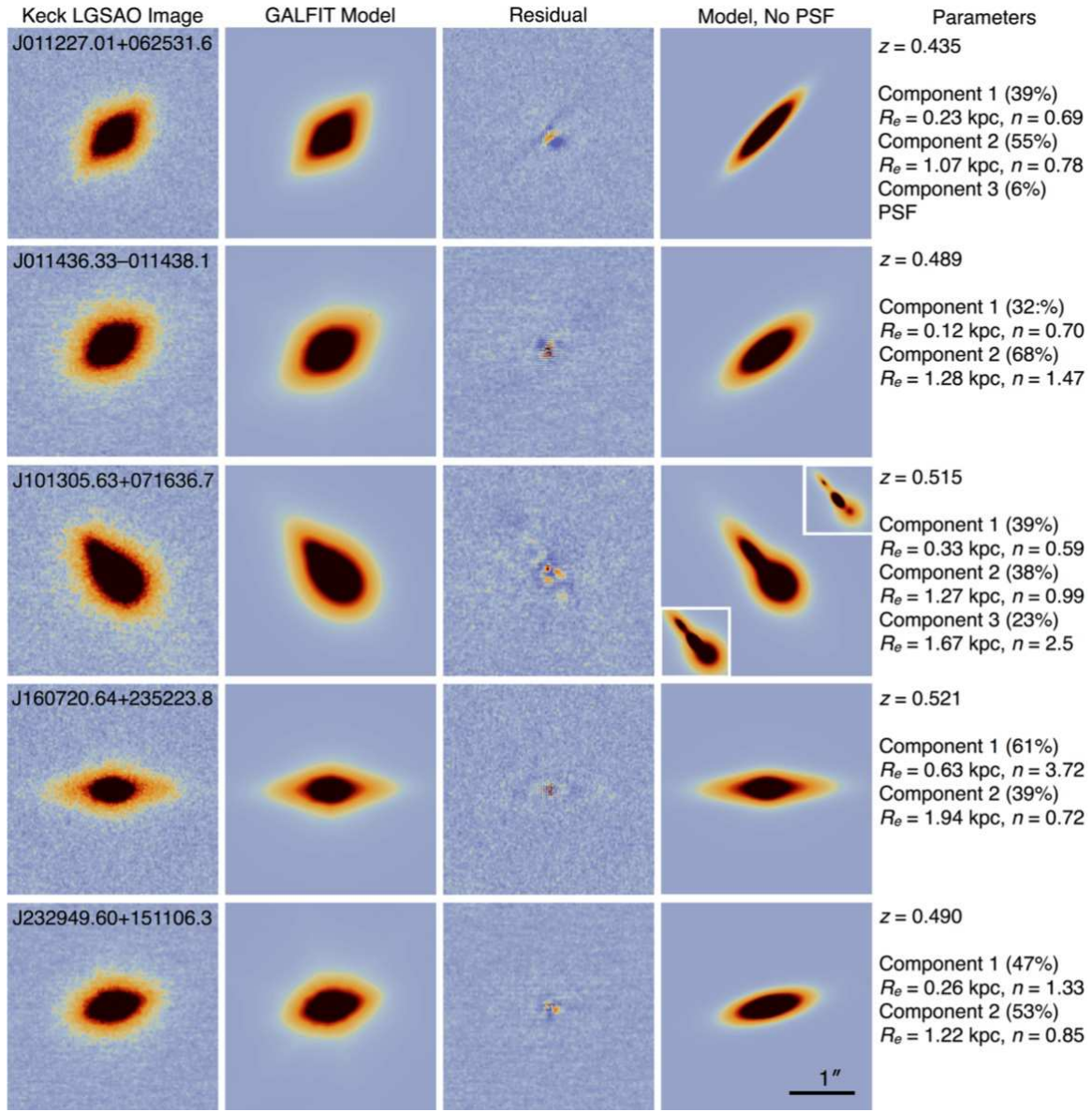}
\caption{Keck LGSAO images and model fits for the 5 galaxies in our sample with single-\sersic\ major-axis effective radii $A_{\rm e} \le 2$ kpc, SED masses $\ge1.5 M_{\odot}$, and initial estimated stellar-population ages $>2$ Gyr. For each galaxy, from left to right, the panels give the observed LGSAO image, the {\sc Galfit} model \citep{pen02,pen10}, the residual from subtracting the model from the data, the {\sc Galfit} model without PSF convolution (which should give the best global impression of the galaxy morphology), and, finally, the spectroscopic redshift and fit parameters for each of the components. Insets for SDSS J101305+071636 give lower-contrast versions to show the three discrete components of this apparently merging system. Each panel is 3\farcs2 square, with N up and E to the left.}\label{morph}
\end{figure*}

We first show in Figure~\ref{morph} the Keck II LGSAO images of the galaxies, along with {\sc Galfit} \citep{pen02,pen10} model fits to these images. 
Note that we initially select these by the major-axis effective radius ($A_{\rm e}$) rather than by the usual circularized effective radius ($R_{\rm e}=A_{\rm e}\sqrt{b/a}$), which we otherwise quote, in order to minimize selection bias against face-on flattened systems. {We select all candidates with $A_{\rm e}\le2$ kpc; nevertheless, because of their small projected axial ratios (a point to which we return in Section \ref{prolate}), all of these so far have ended up with $R_{\rm e}\lesssim1$ kpc, with a maximum of 1.33 kpc.} We have found that we virtually always need to model two components in order to avoid significant systematic residuals. In two special cases (presence of a stellar nucleus; spatially resolved companions), we have had to add a third component. Beside each row we give the parameters derived from the image modeling, together with the spectroscopic redshift.

\begin{figure*}[!t]
\epsscale{1.0}
\plotone{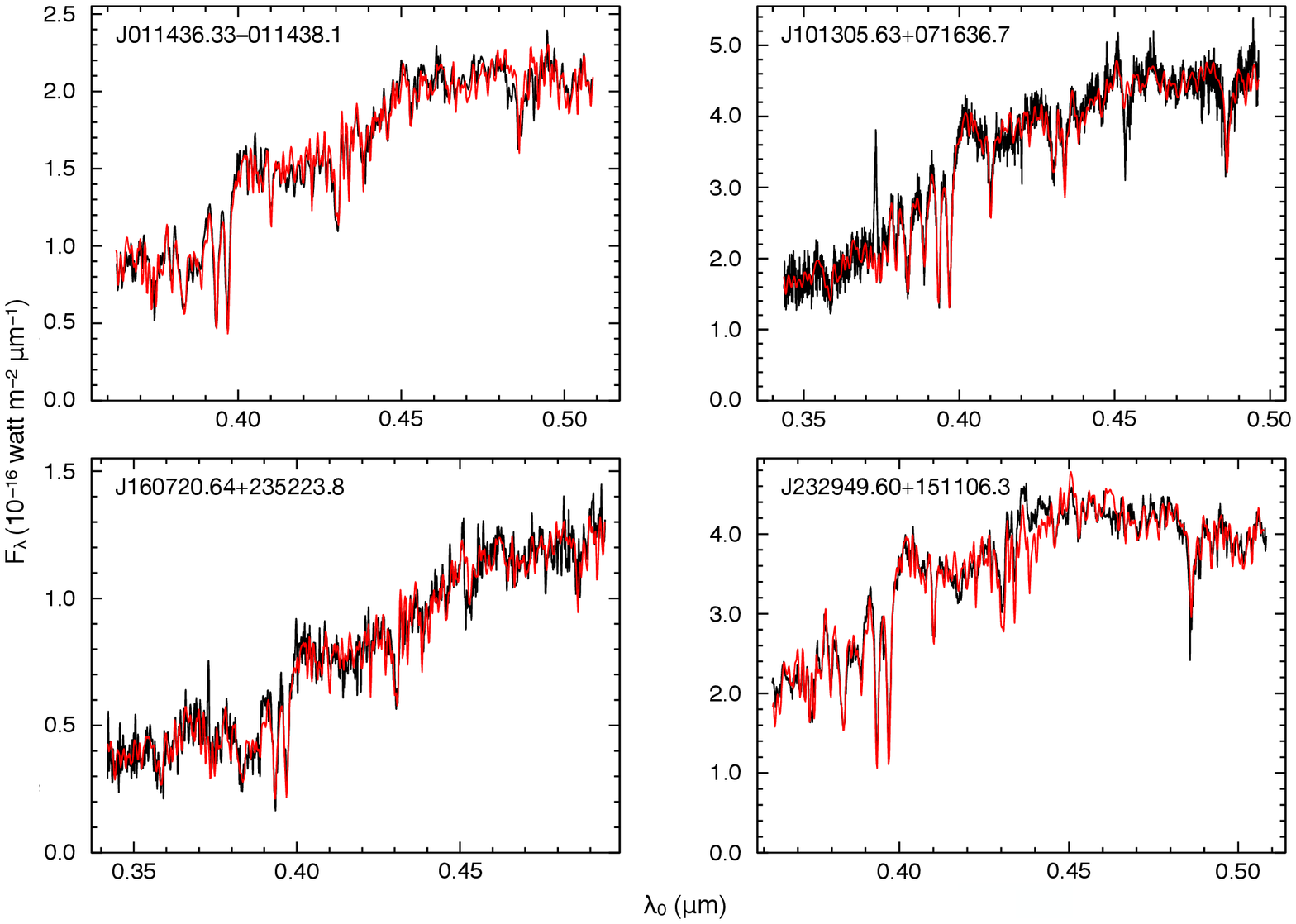}
\caption{pPXF model spectra (red) superposed on observed spectra (black).}\label{spec}
\end{figure*}

Figure~\ref{spec} shows the LRIS spectra for four of the galaxies (for SDSSJ011227 we have only a short Keck II ESI spectrum sufficient to give the redshift). Superposed on the spectra are model fits obtained with the penalized pixel-fitting (pPXF) procedure of \citet{cap04} using the \citet{bru03} spectral synthesis models. SDSSJ101305 and SDSSJ160720 both show some [\ion{O}{2}] $\lambda3727$ emission, which has been excluded from the pPXF fits. 
Since our spectra generally have reliable calibrations, we have used pPXF in the mode for which we fit to both the continuum shape and the absorption lines. The one exception is SDSSJ232949, where our calibration is suspect; in this case, we use a polynomial fit to the continuum; thus the fit is dependent on spectral features alone. Population fitting with pPXF involves choosing a regularization parameter that affects the smoothness of the solutions, and it needs to be emphasized that acceptable fits to the observed spectra can be obtained with a considerable range of star-formation histories. Mostly these involve a simple (but asymmetric) narrowing or widening of the duration of star-forming events, but there can also be modest variations of mass fractions between different events. Given the strong dissipation that must have been involved in the formation of these galaxies, we assume that the initial starburst was likely intense and rather brief. Accordingly, we have chosen regularization parameters that tend to give fairly narrow distributions for the major star-formation episodes, rather than the smoothest possible distribution consistent with the spectra. But it should clearly be understood that this choice is an assumption rather than a result.

\begin{figure*}[p]
\epsscale{1.0}
\plotone{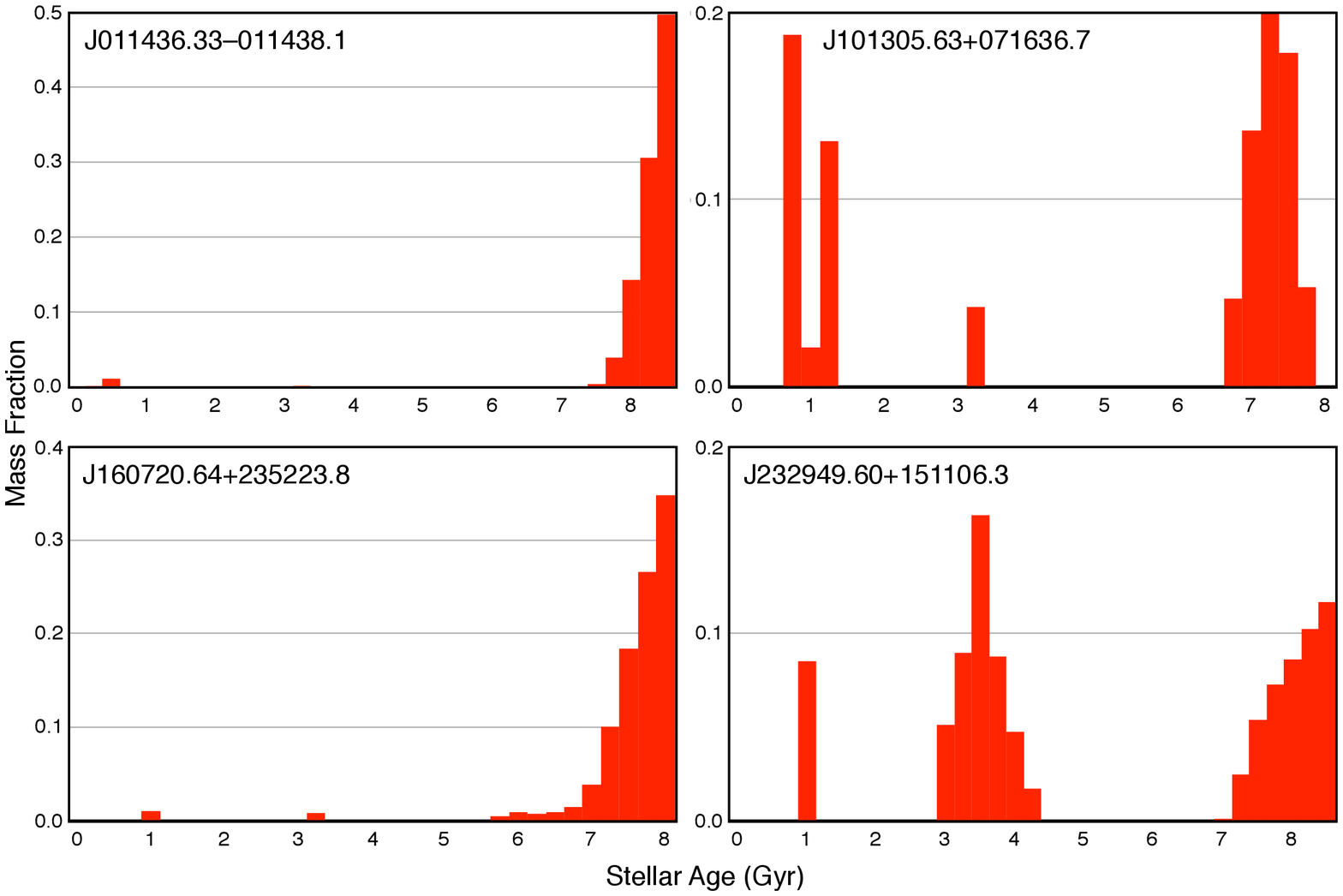}
\caption{Star-formation histories, based on pPXF \citep{cap04} population fitting for four compact passive galaxies. As described in more detail in the text, these should be taken as indicative only, because of degeneracies, particularly in the durations of the star-formation episodes.}\label{ppxf}
\vspace{1cm}
\epsscale{1.0}
\plotone{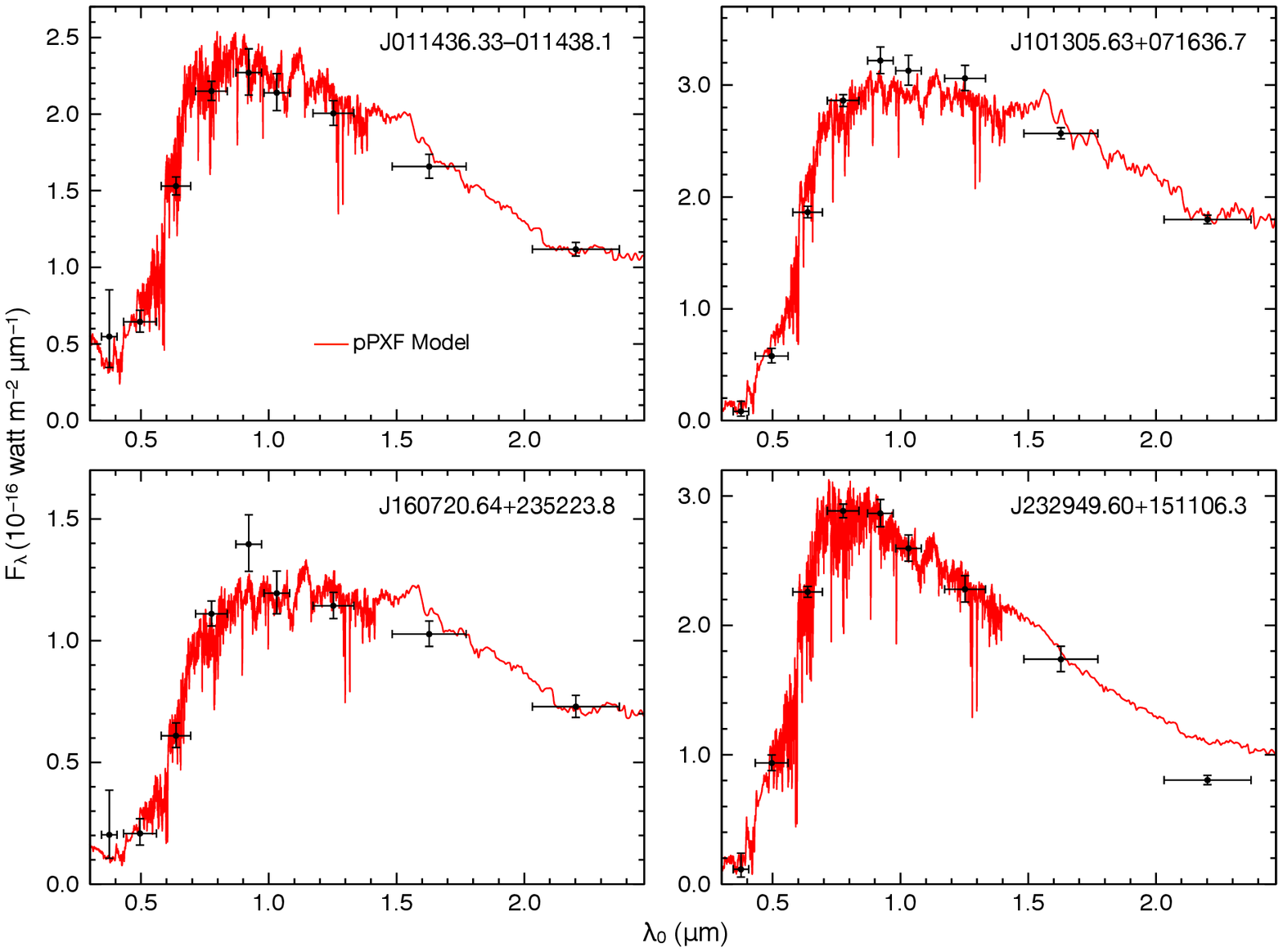}
\caption{SEDs of four compact passive galaxies. The UKIDSS photometry has been slightly scaled as described in the text. The red trace is the pPXF model derived from the spectra.}\label{sed}
\end{figure*}

The star-formation histories corresponding to these models are shown in Figure~\ref{ppxf}, where we have collapsed the three metallicities we have used (0.4, 1.0, and 2.5 solar) to give a clearer indication of the total mass fraction as a function of epoch. The average mass-weighted metallicities of the pPXF-derived populations are 1.4, 2.3, 1.8, and 1.0 solar for SDSSJ011436, SDSSJ101305, SDSSJ160720, and SDSSJ232949, respectively. Even though they are constrained by the reasonably high S/N of our spectra, the fitting uncertainties in these metallicity estimates are typically at least $\sim30$\%, not including any systematic uncertainties that may be present. These metallicities are of course somewhat coupled with the age estimates. 

We can check the consistency of the pPXF star-formation histories, obtained from detailed fitting of both spectral features and the SED over a narrow spectral range (typically 0.35--0.50 $\mu$m, as shown in Figure~\ref{spec}), by plotting the resulting SEDs over the much wider spectral region covered by the SDSS/UKIDSS photometry. These are shown in Figure~\ref{sed}. 
Because the SDSS photometry is based on a model fit to the profile and for the UKIDSS we have used a 2\arcsec-diameter aperture magnitude, we allow for a small offset between them, which will likely depend on the actual source light distribution, among other factors. In order to estimate these offsets, we use Hyper-{\it z} \citep{bol00} with the BC03 models and \citet{cal00} reddening to fit to the SDSS photometry plus the UKIDSS {\it Y} and {\it J} points (only), where the latter magnitudes are given stepped offsets until a minimum $\chi^2$ is found for the best model fit. It should be stressed that because of the small wavelength offsets between the $z$, $Y$, and $J$ points, this determination is not sensitive to the exact SED used; any other consistent fit (\eg\ by ignoring reddening) gives essentially the same offset. The offsets were always in the sense that the UKIDSS flux densities had to be increased. The magnitude offsets for all galaxies except SDSSJ101305 were 0.1 mag or less; for SDSSJ101305 it was 0.35 mag. The offsets were then applied to all of the UKIDSS magnitudes to give the SEDs shown in Fig.~\ref{sed}. These show that the pPXF model fits over their limited spectral region fit the photometry over a much wider region remarkably well, allowing for the occasional errant data point (such as the SDSS $z$-band point for SDSSJ160720).

\clearpage

\section{Survivors from the High-Redshift Compact Passive Galaxy Population?}

\subsection{Stellar Populations}

For the two galaxies discussed in Article 1 (SDSSJ011436 and SDSSJ232949), we concluded from single population fits that the stellar populations of both galaxies had ages of about 5 Gyr, which would have placed their formation epoch at $z\sim1.8$. With the wider options given by the pPXF fitting procedure, we see the possibility that at least some of these galaxies formed most of their mass at high redshifts and therefore may be only slightly modified survivors of the high-redshift population. In particular, SDSSJ011436 is an excellent candidate for such an object, where the pPXF model has $\sim1$\%\ (by mass) of $\sim500$-Myr-old star formation being added to an overwhelmingly dominant population formed at $z>6$. Quite aside from the spectral fits, the photometric SED shown in Figure~\ref{sed} reinforces this picture, where the SDSS $u$-band point, in particular, indicates the effect of the small admixture of a young population. A similar scenario applies to the model for SDSSJ160720: the overwhelming bulk of the mass is formed at high redshift, with a couple of more recent episodes of star formation (perhaps due to wet minor mergers) accounting for $<5$ \%\ of the total mass. 

SDSSJ101305 is more ambiguous, both in terms of star-formation history and in terms of morphology. It appears to comprise three galaxies on the verge of merging, and it is tempting to identify the three discrete episodes of star formation shown in Figure~\ref{ppxf} (at $\sim1$, 3, and 7.5 Gyr) with the three merging components. Whether this is the case or not, there is evidence that a significant portion of the total mass formed at $z\sim5$. Finally, for SDSSJ232949, the pPXF spectral fit (which, as discussed earlier, is dependent only on spectral features, not the continuum shape) indicates major star-formation events at both $z\gtrsim5$ and $z\sim1.2$, with some additional star formation at $z\sim0.65$, or $\sim1$ Gyr before the epoch of observation.

Thus, we end up with two galaxies that appear to have formed 95\%\ or more of their mass at high redshift and two that, while they have significant amounts of mass formed at high redshift, also have considerable mass added at more recent epochs.

\subsection{Masses}

The actual masses of passive compact galaxies are a complex and somewhat contentious subject \citep[see, e.g.,][]{fer12}. We can attempt to estimate the total stellar masses for these galaxies from the adopted stellar population fits, and we can also try to estimate the dynamical masses from the measured velocity dispersions. For the pPXF stellar population fits, we have used BC03 models with \citet{cha03} IMFs. By scaling the mass-fraction-weighted sum of the SSPs contributing to the total spectrum, corrected for reddening, to the photometric SED, we can obtain an estimate of the stellar mass.

We have also used pPXF to fit velocity dispersions to the spectra, following the prescriptions for setting the bias parameter. We can then obtain an estimate of the mass using the virial mass estimator $m_V = \beta R_{\rm e}\sigma^2/G$, where $\beta$ is a parameter that takes into account the relation between $\langle r\rangle$ and $R_{\rm e}$, as well as that between $\langle v^2\rangle$ and $\sigma^2$, where the former in each case are the virial quantities for the stellar distribution and the latter are observable quantities. For a sample of local ellipticals, \citet{cap06} found $\beta=5.0\pm0.1$, and this value has generally been used, with a slight dependence being noted on the \sersic\ parameter $n$ \citep{ber02}. In Table~\ref{mass}, Columns 4 and 5, we compare estimates of the stellar mass derived from our stellar population fits with the dynamical masses, assuming $\beta=5$. For the dynamical masses, we estimate $R_{\rm e}$ from the ellipse enclosing half the total light measured on our {\sc Galfit} model (without PSF convolution). We correct our measured $\sigma$ to the $\sigma$ that would be measured within $R_{\rm e}$ with the relation given by Equation 1 of \cite{cap06}.
\begin{deluxetable*}{lccccc}
\tablecaption{Galaxy Mass Estimates \label{mass}}
\tablewidth{0pt}
\tablecolumns{6}
\tablehead{\colhead{SDSS} & \colhead{$\sigma$} & \colhead{$R_{\rm e}$} & \colhead{Dynam. Mass 1\tablenotemark{a}} & \colhead{pPXF Mass} & \colhead{Dynam. Mass 2\tablenotemark{b}}\\
 \colhead{} & \colhead{(\kms)} & \colhead{(kpc)} & \colhead{($10^{11} M_{\odot}$)} & \colhead{($10^{11} M_{\odot}$)} & \colhead{($10^{11} M_{\odot}$)}}
\startdata
J011436.33$-$011438.1 &$264.4\pm\phn6.5$&0.82&$0.86$&$2.74$&3.50\\
J101305.63+071636.7 &$290.1\pm\phn4.0$&1.33&$1.57$&$2.23$&3.95\\
J160720.64+235223.8 &$268.6\pm\phn8.7$&1.10&$1.13$&$2.14$&3.26\\
J232949.60+151106.3 &$258.8\pm11.0$&0.42&$0.46$&$2.00$&2.80
\enddata
\tablenotetext{a}{Calculated from the virial relation $M_{\rm dyn}=\beta\sigma_{\rm e}^2 R_{\rm e}/G$, with $\beta=5$. The measured velocity dispersions given in column 2 have been corrected to those expected within $R_{\rm e}$ as described by \citet{cap06}.}
\tablenotetext{b}{Calculated from the virial relation, but with $\beta=6(R_{\rm e}/3.185)^{-0.81}(M_*/10^{11})^{0.45}$, following \citet{per13}, where $R_{\rm e}$ is in kpc and $M_*$ is the stellar mass from Column 5, corrected for the difference between our assumed \citet{cha03} IMF and the Salpeter IMF assumed by \citet{per13}, following the prescription of \citet{lon09}.}
\end{deluxetable*}

There are clear discrepancies between these mass estimates, always in the sense that the dynamical mass is less than the mass inferred from the stellar population histories. There are two main assumptions that could explain the difference:

\begin{enumerate}
\item The mass we have derived from the stellar populations assumes the \citet{cha03} initial mass function (IMF). If the actual IMF is more top heavy than we have assumed, these masses could be smaller. However, there is recent evidence that local massive elliptical galaxies may instead have a {\it bottom-heavy} IMF \citep[\eg][]{con12}. Nevertheless, since these compact galaxies will have formed under rather specialized conditions that we do not yet understand very well, we cannot know what sort of IMF would have prevailed, and IMF variation may account for some part of the discrepancy. 
\item In estimating the mass from the velocity dispersion in Column 4, we have assumed that the parameter $\beta$ is the same as that for the local ellipticals studied by \citet{cap06}. This means that we are assuming homology between our compact galaxies and these ellipticals, which is unlikely to be strictly true and may possibly be very far off the mark. The possibility of a substantial difference in structure is reinforced by the recent work of \citet{per13}, who find an approximate empirical relation between $\beta$ ($K$ in their paper) and the compactness of a stellar system at a given mass. When we apply their correction, we get the dynamical masses given in the last column of Table~\ref{mass}. 
These masses are all somewhat above our estimate of the stellar mass, possibly because of contributions from dark matter. However, given the simplifications and uncertainties involved, these values probably are not significantly discrepant.
\end{enumerate}

\subsection{Prolate Morphologies?}\label{prolate}

The models shown in the last column of Figure~\ref{morph} have $b/a$ ratios for single \sersic\ fits ranging from 0.26 to 0.47 (for SDSSJ101305, we consider only the central component; if we were to include all components, the ratio would be even less). These low ratios are similar to those often found for passive galaxies at high redshift \citep{sto04,sto07,vdwel11}. While our sample is too small to be more than indicative of a possibility, it does serve as a basis for speculating that the intrinsic morphologies of many of these galaxies, at both high and low redshifts, may in fact be prolate. One can easily show that the distribution of projected axial ratios of an ensemble of disk galaxies with a given intrinsic $b/a$ will be approximately uniform over the range from the intrinsic value to 1, whereas the distribution for prolate objects will be strongly peaked near the intrinsic ratio. For example, formally, with the probability for a randomly oriented thin disk to have a projected $b/a \le 0.5$ being 1/3, finding five such objects all with projected $b/a \le 0.5$ has a probability of only 0.4\%. If, on the other hand, we consider an idealized rod with intrinsic $b/a=0.2$ (one cannot use an infinitely thin rod for obvious reasons), the probability of having five objects with projected $b/a\le0.5$ is about 22\%. Since our targets were initially selected from the SDSS/UKIDSS surveys, where they all look essentially stellar, and since our final selection was based on the semimajor axis parameter $A_{\rm e}$, there should be little or no selection bias favoring small axial ratios. The possibility that passive compact galaxies at high redshift might be prolate was briefly considered and rejected by \citet{vdwel11}, on the grounds that such objects are not seen locally and that galaxy formation mechanisms should ``be independent of cosmic time.'' We believe, on the contrary, that while we have no compelling model for how these massive compact galaxies form, the very fact that passive galaxies in the early universe typically have morphologies quite unlike those seen locally is an indication that galaxy formation mechanisms have evolved over cosmic time. If these galaxies were actually to be found to be prolate, such a morphology would imply an anisotropic stellar velocity field and structural differences that could very well account for the discrepancy between the calculated stellar mass and the dynamical mass estimated from the relation that works well for the population of local ellipticals.

\section{Overview}

{Our search for examples of galaxies with $M_*>10^{11} M_{\odot}$, $R_{\rm e} \lesssim 1$ kpc at $z\sim0.5$, with the overwhelming bulk of the stellar mass having been formed at $z>5$, is by no means complete, although it does seem to indicate that such galaxies are quite rare at this redshift. Nevertheless, our stellar population analysis with pPXF, having found at least two galaxies that seem to fit this description, does seem to indicate that it is possible to identify a few galaxies at this redshift that come very close to being unmodified survivors from the high-redshift population.} In such cases, it would often be extremely valuable to obtain UV spectra in order to place firmer constraints on the contaminating young population. While mass determinations are still uncertain, the arguments of \citet{per13} on the likely lack of homology with the local elliptical galaxy population, as well as the possibility that the stellar velocity distribution may be highly anisotropic, lead us, like \citet{fer12}, to tend to trust the stellar population masses over the dynamical mass estimates. If the stellar mass estimates are correct, our galaxies typically have masses of 2--$3\times10^{11} M_{\odot}$. Although our current sample is small, the fact that all five of our galaxies have small $b/a$ ratios has led us to tentatively consider the possibility that these galaxies may have prolate morphologies. It will be necessary to identify a larger sample to test this suggestion. 

As indicated in Section 1, we still do not seem to have a wholly adequate understanding of the nature of the apparently highly dissipative process by which these massive compact galaxies form. If prolate morphologies were to be confirmed for a significant subset of these galaxies, this finding would almost certainly help point the way to a better understanding of their formation mechanism.

After this article was submitted, \citet{tru13} announced the remarkable discovery that NGC1277, at a distance of only $\sim73$ Mpc, is an apparently essentially unaltered survivor from the high-redshift massive compact galaxy population. This discovery gives us some hope that it may be possible to identify a reasonable sample of such galaxies at fairly low redshifts.

\acknowledgments
We thank Luis Peralta de Arriba for pointing out an error in the original version of Table 1 and for other helpful comments. 
This research has been partially supported by NSF grant AST-0807900. The UKIDSS project is defined in \citet{law07}. UKIDSS uses the UKIRT Wide Field Camera (WFCAM; \citealt{cas07}) and a photometric system described in \citet{hew06}. The pipeline processing and science archive are described in \citet{ham08}. Funding for the SDSS has been provided by the Alfred P. Sloan Foundation, the Participating Institutions, the National Science Foundation, the U.S. Department of Energy, the National Aeronautics and Space Administration, the Japanese Monbukagakusho, the Max Planck Society, and the Higher Education Funding Council for England. The SDSS website is found at http://www.sdss.org/.

{\it Facilities:} \facility{Keck:I (LRIS)}, \facility{Keck:II (LGSAO/NIRC2)}, \facility{Sloan (SDSS)}, \facility{UKIRT (UKIDSS)}




\begin{thebibliography}
\expandafter\ifx\csname natexlab\endcsname\relax\def\natexlab#1{#1}\fi

\bibitem[Bertin \etal(2002)]{ber02} Bertin, G., Ciotti, L., \& Del Principe, M. 2002, \aap, 386, 149

\bibitem[{Bezanson {et~al.}(2009)Bezanson, van Dokkum, Tal, Marchesini, Kriek,
  Franx, \& Coppi}]{bez09}
Bezanson, R., van Dokkum, P.~G., Tal, T., \etal\ 2009, \apjl, 697, 1290

\bibitem[{Bolzonella {et~al.}(2000)Bolzonella, Miralles, \& Pell{\'o}}]{bol00}
Bolzonella, M., Miralles, J.~M., \& Pell{\'o}, R. 2000, \aap, 363, 476

\bibitem[Bruzual \& Charlot(2003)]{bru03} Bruzual A., G., \& Charlot, S. 2003, \mnras, 344, 1000

\bibitem[{Buitrago \etal(2008)}]{bui08} Buitrago, F., Trujillo, I., Conselice, C. J., \etal\ 2008, \apjl, 687, L61

\bibitem[Calvi \etal(2011)]{cal11} Calvi, R., Poggianti, B. M., \& Vulcani, B. 2011, \mnras, 416, 727

\bibitem[{Calzetti {et~al.}(2000)Calzetti, Armus, Bohlin, Kinney, Koornneef, \&
  Storchi-Bergmann}]{cal00}
Calzetti, D., Armus, L., Bohlin, R.~C., \etal\ 2000, \apj, 533, 682
  
\bibitem[Cappellari \etal(2006)]{cap06} Cappellari, M., Bacon, R., Bureau, M., \etal\ 2006, \mnras, 366, 1126

\bibitem[Cappellari \& Emsellem(2004)]{cap04} Cappellari, M., \& Emsellem, E. 2004, \pasp, 115, 138

\bibitem[{Casali {et~al.}(2007)Casali, Adamson, Alves~de Oliveira, Almaini,
  Burch, Chuter, Elliot, Folger, Foucaud, Hambly, Hastie, Henry, Hirst, Irwin,
  Ives, Lawrence, Laidlaw, Lee, Lewis, Lunney, McLay, Montgomery, Pickup, Read,
  Rees, Robson, Sekiguchi, Vick, Warren, \& Woodward}]{cas07}
Casali, M., Adamson, A., Alves~de Oliveiera, A., {et~al.} 2007, \aap, 467, 777

\bibitem[{Cassata \etal(2010)}]{cas10} Cassata, P., Giavalisco, M., Guo, Y., \etal\ 2010, \apjl, 714, L79

\bibitem[{Chabrier(2003)}]{cha03}
Chabrier, G. 2003, \pasp,
  115, 763

\bibitem[Conroy \& van Dokkum(2012)]{con12} Conroy, C. \& van Dokkum, P. G. 2012, \apj, 760, 71.

\bibitem[{Daddi \etal(2005)}]{dad05} Daddi, E., Renzini, A., Pirzkal, N., \etal\ 2005, \apj, 626, 680

\bibitem[Damjanov \etal(2013)]{dam13} Damjanov, I., Chilingarian, I., Hwang, H. S., \& Geller, M. J. 2013, \apjl, 775, L48

\bibitem[{Damjanov {et~al.}(2009)Damjanov, McCarthy, Abraham, Glazebrook, Yan,
  Mentuch, LeBorgne, Savaglio, Crampton, Murowinski, Juneau, Carlberg,
  J{\o}rgensen, Roth, Chen, \& Marzke}]{dam09}
Damjanov, I., McCarthy, P. J., Abraham, R. G., {et~al.} 2009, \apjl, 695, 101

\bibitem[Ferr\'{e}-Mateu \etal(2012)]{fer12} Ferr\'{e}-Mateu, A., Vazdekis, A., Trujillo, I., \etal\ 2012, \mnras, 423, 632
  
\bibitem[{Hambly {et~al.}(2008)Hambly, Collins, Cross, Mann, Read, Sutorius,
  Bond, Bryant, Emerson, Lawrence, Rimoldini, Stewart, Williams, Adamson,
  Hirst, Dye, \& Warren}]{ham08}
Hambly, N.~C., Collins. R. S., Cross, N. J. G., {et~al.} 2008, \mnras, 384, 637

\bibitem[{Hewett {et~al.}(2006)Hewett, Warren, Leggett, \& Hodgkin}]{hew06}
Hewett, P.~C., Warren, S.~J., Leggett, S.~K., \& Hodgkin, S.~T. 2006, \mnras, 367, 454
  
\bibitem[{Hopkins {et~al.}(2009)Hopkins, Bundy, Murray, Quataert, Lauer, \&
  Ma}]{hop09}
Hopkins, P.~F., Bundy, K., Murray, N., \etal\ 
  2009, \mnras, 398, 898

\bibitem[{Kriek \etal(2008)}]{kri08} Kriek, M., van der Wel, A., van Dokkum, P. G., Franx, M., \& Illingworth, G. D. 2008, \apj, 682, 896

\bibitem[{Kriek \etal(2006)}]{kri06} Kriek, M., van Dokkum, P. G., Franx, M., \etal\ 2006,  \apjl, 649, L71

\bibitem[{Lawrence {et~al.}(2007)Lawrence, Warren, Almaini, Edge, Hambly,
  Jameson, Lucas, Casali, Adamson, Dye, Emerson, Foucaud, Hewett, Hirst,
  Hodgkin, Irwin, Lodieu, McMahon, Simpson, Smail, Mortlock, \& Folger}]{law07}
Lawrence, A., Warren, S. J., Almaini, O., {et~al.} 2007, \mnras,
  379, 1599

\bibitem[Longhetti \& Saracco(2009)]{lon09} Longhetti, M., \& Saracco, P. 2009, \mnras, 394, 774
  
\bibitem[{McGrath \etal(2008)}]{mcg08} McGrath, E. J., Stockton, A., Canalizo, G., Iye, M., \& Maihara, T. 2008, \apj, 682, 303

\bibitem[{Muzzin \etal(2009)}]{muz09} Muzzin, A., van Dokkum, P., Franx, M., \etal\ 2009, \apjl, 706, L188

\bibitem[{Oke {et~al.}(1995)Oke, Cohen, Carr, Cromer, Dingizian, Harris,
  Labrecque, Lucinio, Schaal, Epps, \& Miller}]{oke95}
Oke, J.~B., Cohen, J. G., Carr, M., {et~al.} 1995, \pasp, 107, 375

\bibitem[Oogi \& Habe(2013)]{oogi13} Oogi, T., \& Habe, A. 2013, \mnras, 657, 641

\bibitem[Peng \etal(2002)Peng, Ho, Impey, \& Rix]{pen02}
Peng, C.~Y., Ho, L.~C., Impey, C.~D., \& Rix, H.-W. 2002, \aj,
  124, 266
  
\bibitem[Peng \etal(2010a)]{pen10} Peng, C.~Y., Ho, L.~C., Impey, C.~D., \& Rix, H.-W. 2010a, \aj, 139, 2097

\bibitem[Peng \etal(2010b)]{ypen10} Peng, Y.-J., Lilly, S. J., Kovac, K.,\etal\ 2010b, \apj, 721, 193

\bibitem[Peralta de Arriba \etal(2013)]{per13} Peralta de Arriba, L., Balcells, M., Falc\'{o}n-Barroso, J., \& Trujillo, I. 2013, arXiv:1307.4376

\bibitem[Poggianti \etal(2013)]{pog13} Poggianti, B. M., Calvi, R., Bindoni, D., \etal\ 2013, \apj, 762, 77

\bibitem[Quilis \& Trujillo(2013)]{qui13} Quilis, V., \& Trujillo, I. 2013, \apjl, 773, L8

\bibitem[{Shih \& Stockton(2011)}]{shi11} Shih, H.-Y., \& Stockton, A. 2011, \apj, 733, 45

\bibitem[{Sommer-Larsen \& Toft(2010)}]{som10} Sommer-Larsen, J., \& Toft, S. 2010, \apj, 721, 1755

\bibitem[{Stockton {et~al.}(2004)Stockton, Canalizo, \& Maihara}]{sto04}
Stockton, A., Canalizo, G., \& Maihara, T. 2004, \apj, 605, 37

\bibitem[{Stockton \& McGrath(2007)}]{sto07}
Stockton, A., \& McGrath, E. 2007, in ASP Conf. Ser. 379, ed. N.~Metcalfe \&
  T.~Shanks, (San Francisco, CA: ASP), 122

\bibitem[{Stockton \etal(2010)Stockton, Shih, \& Larson}]{sto10} Stockton, A., Shih, H.-Y., \& Larson, K. 2010, \apjl, 709, L58

\bibitem[{Taylor {et~al.}(2009)Taylor, Franx, Glazebrook, Brinchmann, van~der
  Wel, \& van Dokkum}]{tay09}
Taylor, E.~N., Franx, M., Glazebrook, K., \etal\ 2009, \apj, 720, 723

\bibitem[{Toft \etal(2012)}]{tof12} Toft, S., Gallazzi, A., Zirm, A., \etal\ 2012, \apj, 754, 3

\bibitem[{Toft \etal(2007)}]{tof07} Toft, S., van Dokkum, P., Franx, M., \etal\ 2007, \apj, 671, 285

\bibitem[{Trujillo \etal(2009)}]{tru09} Trujillo, I., Cenarro, A. J., de Lorenzo-C\'{a}ceres, A., \etal\ 2009, \apjl, 692, L118

\bibitem[{Trujillo \etal(2007)}]{tru07} Trujillo, I., Conselice, C. J., Bundy, K., \etal\ 2007, \mnras, 382, 109

\bibitem[Trujillo \etal(2013)]{tru13} Trujillo, I., Ferr\'{e}-Mateu, Balcells, M., Vazdekis, A., \& S\'{a}nchez-Bl\'{a}zquez 2013, arXiv:1310.6367

\bibitem[{Valentinuzzi {et~al.}(2010)Valentinuzzi, Fritz, Poggianti, Bettoni,
  Cava, Fasano, D'Onofrio, Couch, Dressler, Moles, Moretti, Omizzolo,
  Kjaergaard, Vanzella, \& Varela}]{val10}
Valentinuzzi, T., Fritz, J., Poggianti, B. M., {et~al.} 2010, \apj, 712, 226

\bibitem[{van der Wel \etal(2011)}]{vdwel11} van der Wel, Rix, H.-W., Wuyts, S., \etal\ 2011, \apj, 730, 38

\bibitem[van de Sande \etal(2011)]{vdS11} van de Sande, J., Kriek, M., Franx, M., \etal\ 2011, \apjl, 736, L9

\bibitem[van de Sande \etal(2013)]{vdS13} van de Sande, J., Kriek, M., Franx, M., \etal\ 2013, \apj, 771, 85

\bibitem[{van Dokkum {et~al.}(2008)van Dokkum, Franx, Kriek, Holden,
  Illingworth, Magee, Bouwens, Marchesini, Quadri, Rudnick, Taylor, \&
  Toft}]{vdok08}
van Dokkum, P.~G., Franx, M., Kriek, M., {et~al.} 2008, \apjl, 677, L5

\bibitem[{van Dokkum {et~al.}(2009)van Dokkum, Kriek, \& Franx}]{vdok09}
van Dokkum, P.~G., Kriek, M., \& Franx, M. 2009, \nat, 460, 717

\bibitem[{Wizinowich {et~al.}(2006)Wizinowich, Le~Mignant, Bouchez, Campbell,
  Chin, Contos, van Dam, Hartman, Johansson, Lafon, Lewis, Stomski, Summers,
  Brown, Danforth, Max, \& Pennington}]{wiz06}
Wizinowich, P.~L., Le Mignant, D., Bouche, A. H., {et~al.} 2006, \pasp, 118, 297

\bibitem[{Wuyts \etal(2010)}]{wuy10} Wuyts, S, Cox, T., Hayward, C. C., \etal\ 2010, \apj, 722, 1666

\bibitem[Zibetti \etal(2012)]{zib12} Zibetti, S., Gallazzi, A., Charlot, S., Pierini, D., \& Pasquali, A. 2012, \mnras, 428, 1479

\bibitem[{Zirm \etal(2007)}]{zir07} Zirm, A. W., van der Wel, A., Franx, M., \etal\ 2007, \apj, 656, 66

\end{thebibliography}
\end{document}